# Electronic structures of layered perovskite Sr$_2$MO$_4$ (M=Ru, Rh, and Ir)

S. J. Moon,[1] M. W. Kim,[1] K. W. Kim,[1] Y. S. Lee,[2] J.-Y. Kim,[3] J.-H. Park,[3,4] B. J. Kim,[5] S.-J. Oh,[5] S. Nakatsuji,[6] Y. Maeno,[6] I. Nagai,[7] S. I. Ikeda,[7] G. Cao,[8] and T. W. Noh[1,*]

[1]*ReCOE & FPRD, School of Physics and Astronomy, Seoul National University, Seoul 151-747, Korea*

[2]*Department of Physics, Soongsil University, Seoul 156-743, Korea*

[3]*Pohang Accelerator Laboratory, Postech, Pohang 790-784, Korea*

[4]*Department of Physics & electron Spin Science Center, Postech, Pohang 790-784, Korea*

[5]*CSCMR & FPRD, School of Physics and Astronomy, Seoul National University, Seoul 151-747, Korea*

[6]*Department of Physics, Kyoto University, Kyoto 606-8501, Japan*

[7]*National Institute of Advanced Industrial Science and Technology, Tsukuba, Ibaraki 305-8568, Japan*

[8]*Department of Physics and Astronomy, University of Kentucky, Lexington, Kentucky 40506, USA*

**Abstract**

We investigated the electronic structures of the two-dimensional layered perovskite Sr$_2$MO$_4$ (M=4d Ru, 4d Rh, and 5d Ir) using optical spectroscopy and polarization-dependent O 1s x-ray absorption spectroscopy. While the ground states of the series of compounds are rather different, their optical conductivity spectra $\sigma(\omega)$ exhibit similar interband transitions, indicative of the common electronic structures of the 4d and 5d layered oxides. The energy splittings between the two $e_g$ orbitals, *i.e.*, $d\,3z^2\text{-}r^2$ and $d\,x^2\text{-}y^2$, are about 2 eV, which is much larger than those in the pseudocubic and 3d layered perovskite oxides. The electronic properties of the Sr$_2$MO$_4$ compounds are discussed in terms of the crystal structure and the extended character of the 4d and 5d orbitals.

---

[*]E-mail address: twnoh@snu.ac.kr

Transition metal oxides (TMO) with layered perovskite structures have attracted much attention due to their exotic physical properties, such as high $T_c$ superconductivity in cuprates,[1] one-dimensional charge/spin self-organization in nickelates,[2] and charge/orbital ordering in manganites.[3] These phenomena in 3$d$ TMO originate from the low dimensionality and the related anisotropy in the electronic structure, as well as generic correlation effects in the $d$ orbitals. The discovery of unconventional superconductivity in 4$d$ layered $Sr_2RuO_4$ has initiated intense interest in 4$d$ and 5$d$ layered compounds because of the potential to discover novel phenomena.[4]

The electronic structures of 4$d$ and 5$d$ layered TMO should be quite different from those of 3$d$ TMO. For a 3$d$ TMO, the Hund coupling energy $J_H$ (~3 eV) is much larger than the crystal field splitting 10$Dq$ (1–2 eV) between the $t_{2g}$ and $e_g$ states.[5,6] For example, the electronic structure of $LaSrMnO_4$ (a 3$d^4$ system) with a high spin configuration is drawn schematically in Fig. 1(a).[5,6] The exchange splitting of the $e_g$ states is sufficiently large to place the $e_g^{\uparrow}$ band below the $t_{2g}^{\downarrow}$ band. The elongation of $MnO_6$ octahedra results in energy splitting between the two $e_g$ states (*i.e.*, $d\ 3z^2$-$r^2$ and $d\ x^2$-$y^2$), and this value is typically about 1 eV. Since most interband transitions are very broad, this $e_g$ orbital splitting is barely observed in the optical conductivity spectra $\sigma(\omega)$. In contrast, for 4$d$ and 5$d$ TMO, the crystal field splitting should be larger than the exchange splitting due to the extended nature of the 4$d$ and 5$d$ orbitals. Therefore, they are usually in the low spin configuration, in which all of the $t_{2g}$ bands are lower than the $e_g$ bands. According to recent x-ray absorption spectroscopy (XAS) studies on the Ca$_{2-x}$Sr$_x$RuO$_4$ system, the energy splitting between $d\ 3z^2\text{-}r^2$ and $d\ x^2\text{-}y^2$ is sizable, while the exchange splitting in the $e_g$ state is not detectable, as shown in Fig. 1(b).[7] [The $e_g$ orbital splitting between $d\ 3z^2\text{-}r^2$ and $d\ x^2\text{-}y^2$ was reported to be about 3 eV, which is quite unusual. To the best of our knowledge, large $e_g$ orbital splitting has not been reported in any other compound.]

To attain further insights into the electronic structure of 4$d$ and 5$d$ layered compounds, we made systematic studies of Sr$_2M$O$_4$ ($M$=Ru, Rh, and Ir). Each Sr$_2M$O$_4$ has the layered perovskite structure. The space group symmetry of Sr$_2$RuO$_4$ is $I4/mmm$, and the Ru-O-Ru bond angle is 180°.[8] In Sr$_2$RhO$_4$ and Sr$_2$IrO$_4$, the $M$O$_6$ octahedra are rotated with respect to the $c$-axis, so their space group symmetry becomes $I4_1/acd$.[9,10] The $d$-orbital configurations of the Ru$^{4+}$, Rh$^{4+}$, and Ir$^{4+}$ ions are 4$d^4$, 4$d^5$, and 5$d^5$, respectively. Despite having a similar crystal structure, their electronic ground states differ somewhat. While Sr$_2$RuO$_4$ is a superconductor below about 1 K, the paramagnetic metallic state of Sr$_2$RhO$_4$ is retained to 36 mK.[11] Interestingly, Sr$_2$IrO$_4$ is an insulator with a canted ferromagnetic ordering.[10]

In this paper, we report the optical conductivity spectra $\sigma(\omega)$ and polarization-dependent O 1$s$ XAS data for Sr$_2M$O$_4$ ($M$=Ru, Rh, and Ir). The optical spectra showed three charge transfer transitions in the energy region between 0 and 8 eV with a systematic trend with $M$. By comparing $\sigma(\omega)$ with the XAS spectra, we could determine the electronic structures of Sr$_2M$O$_4$. In particular, we found that the splitting of the $e_g$ orbitals due to the elongation of the $M$O$_6$ octahedra along the $c$-axis is quite large, about 2 eV. We also discuss the low energy optical responses of Sr$_2M$O$_4$ in relation to their ground states.

Single crystals of Sr$_2$RuO$_4$ and Sr$_2$RhO$_4$ were grown using the floating zone method,[4,11] and the Sr$_2$IrO$_4$ single crystalline sample was grown using the flux technique.[10] The magnetic and transport properties of Sr$_2$RhO$_{4-\delta}$ depend on its oxygen contents.[11] We used stoichiometric Sr$_2$RhO$_{4.00}$. We measured the $ab$-plane reflectivity spectra $R(\omega)$ at room temperature over a wide photon energy region between 5 meV and 30 eV. The corresponding $\sigma(\omega)$ were obtained using the Kramers–Kronig (KK) transformation of $R(\omega)$. We checked the validity of our KK analysis by measuring $\sigma(\omega)$ between 0.6 and 6.4 eV independently using spectroscopic ellipsometry.

Figure 2 shows the $ab$-plane $\sigma(\omega)$ of Sr$_2$RuO$_4$, Sr$_2$RhO$_4$, and Sr$_2$IrO$_4$. First, we will focus on the optical transitions above 1.5 eV. As shown in Fig. 2, the $\sigma(\omega)$ of Sr$_2$RuO$_4$, Sr$_2$RhO$_4$, and Sr$_2$IrO$_4$ all showed three interband transitions. According to the literature on regular perovskite TMO, the optical transitions from the O 2$p$ to Sr $d$ bands are usually located at about 9 to 10 eV.[13,14] Therefore, all of the A, B, and C peaks observed in Fig. 2 might come from the charge transfer transition from O 2$p$ to M $d$ bands.

Polarization-dependent O 1$s$ XAS spectra of Sr$_2$RuO$_4$ have already been reported. To obtain further insights, we measured O 1$s$ XAS spectra of Sr$_2$RhO$_4$ and Sr$_2$IrO$_4$ at the EPU6 beamline of the Pohang Light Source (PLS). O 1$s$ XAS detects the transition from O 1$s$ to O 2$p$ orbitals. The XAS measurements can probe the unoccupied density of states that are strongly hybridized with the O 2$p$ orbitals. With the incident $E$ vector mainly in the in-plane (*i.e.*, $\theta=0°$), the XAS spectra should show only the M $d$ orbital states that are strongly coupled with O 2$p_{x/y}$. With $\theta=60°$, the XAS spectra should show the M $d$ orbital states coupled with both O 2$p_{x/y}$ and O 2$p_z$, but the latter states should make a larger contribution. The polarization dependence should be large, since the $d$ orbitals have strong directional dependence. The main bondings of the in-plane O 2$p_{x/y}$ orbitals are O 2$p_{x/y}$-$d\ xy$ ($xy_P$) and O 2$p_{x/y}$-$d\ x^2$-$y^2$ ($x^2$-$y^2_P$), and that of the apical O 2$p_{x/y}$ orbitals is O 2$p_{x/y}$-$d\ yz/zx$ ($yz/zx_A$), while those of the in-plane and apical O 2$p_z$ orbitals are O 2$p_z$-$d\ yz/zx$ ($yz/zx_P$) and O 2$p_z$-$d\ 3z^2$-$r^2$ ($z^2_A$), respectively. The $d\ 3z^2$-$r^2$ orbital mainly bonds with the apical O $2p_z$, but still bonds weakly with the in-plane O $2p_{x/y}$ ($z^2_P$). Therefore, the spectra with $\theta=0°$ should show the contributions of the $xy_P$, $x^2-y^2_P$, $yz/zx_A$, and $z^2_P$ states, while the $yz/zx_P$ and $z^2_A$ states should contribute more strongly in the spectra with $\theta=60°$.

Figure 3 shows the polarization-dependent XAS spectra of $Sr_2MO_4$. [The XAS spectra of $Sr_2RuO_4$ (Ref. 7) is included in Fig. 3(a) for better comparison with those of other compounds.] All three compounds exhibit very similar XAS spectra with strong polarization dependence. Considering the directional dependence of the $p$-$d$ hybridization, the four peaks in the $\theta=0°$ spectra can be assigned as $yz/zx_A$, $xy_P$, $z^2_P$, and $x^2-y^2_P$, and the two peaks in the $\theta=60°$ spectra can be assigned as $yz/zx_P$ and $z^2_A$ from the lowest energy. In the layered TMO, the core-hole energy of the apical oxygen is lower than that of the in-plane oxygen due to the difference in the chemical environment.[7] Therefore, the $t_{2g}$ states related to the apical oxygen ($yz/zx_A$) should have lower energy than those related to the in-plane oxygen ($xy_P$), which is also the case for the $e_g$ orbitals ($z^2_A$ and $z^2_P$).

These peak assignments in the XAS spectra are also consistent with those of $\sigma(\omega)$ obtained from the optical measurements. From the XAS results, three $p$-$d$ charge transfer transitions are predicted to be observed in the $\sigma(\omega)$: those from O $2p$ to $d$ *xy/yz/zx* (peak A), *d 3z²-r²* (peak B), and *d x²-y²* (peak C) from the lowest energy. This is well reproduced in $\sigma(\omega)$, as shown in Fig. 2. Using the Lorentz oscillators, we estimated the peak positions of the interband transitions to be about 3.7, 4.7, and 6.7 eV for $Sr_2RuO_4$, about 2.2, 3.6, and 6.3 eV for $Sr_2RhO_4$, and about 3.3, 5.4, and 7.5 eV for $Sr_2IrO_4$. We summarized the peak positions in $\sigma(\omega)$ and the XAS spectra in Fig. 4. It is clear that the two independently determined peak positions are quite consistent, which indicates the self-consistency of our assignments.

It is interesting to examine the change in the charge transfer energies in the $Sr_2MO_4$ series with *M*. As shown in Fig. 4, the charge transfer peaks of $Sr_2RhO_4$ have the lowest values in the series. This change with *M* could be explained by the ionic model, which has been applied to explain the electronic structures of numerous 3*d* TMO.[15,16] According to the ionic model, the charge transfer energy decreases as the ionization energy of a transition metal increases. For a given oxidation state, the ionization energy of a transition metal increases with the atomic number and decreasing principal quantum number. Therefore, the charge transfer energies of $Sr_2RhO_4$ should be the smallest in our $Sr_2MO_4$ series. In this respect, the systematic changes in the charge transfer energies in our 4*d* and 5*d* layered oxides can be explained in terms of the ionic model.

It is unexpected that the energy splittings of the $d\,3z^2\text{-}r^2$ and $d\,x^2\text{-}y^2$ orbitals for Sr$_2$RhO$_4$ and Sr$_2$IrO$_4$ are quite large, *i.e.*, ~2 eV, like Sr$_2$RuO$_4$. These $e_g$ orbital splittings have not been observed in pseudocubic perovskite $4d$ Sr$M$O$_3$, which have nearly undistorted $M$O$_6$ octahedra.[14] For the layered Sr$_2M$O$_4$ compounds, the $M$O$_6$ octahedra is elongated along the *c*-axis.[8–10] It is believed that $t_{2g}$ and the $e_g$ orbitals split in such a way that the energy of the $d\,xy$ and $d\,x^2\text{-}y^2$ orbitals is higher than that of $d\,yz/zx$ and $d\,3z^2\text{-}r^2$, respectively. This simple idea appears to agree with our finding that the $e_g$ orbital splitting becomes rather large (~2 eV), while that of the $t_{2g}$ states is negligible. Indeed, the energy splitting of the $t_{2g}$ states in Ca$_2$RuO$_4$ should be only 0.2 eV.[17,18] Since the hybridization of the O $2p$ orbitals should be stronger with the $e_g$ than with the $t_{2g}$ orbitals, the energy states of the $e_g$ orbitals could be more sensitive to lattice distortion, which can lead to larger $e_g$ orbital splitting. Further investigations of the large $e_g$ orbital splitting in other $4d$ and $5d$ layered materials are warranted.

Compared to our $4d$ and $5d$ layered compounds, the reported values of the $e_g$ orbital splittings for $3d$ layered TMO are much smaller. XAS measurement of the layered manganite La$_{2-x}$Sr$_{1+2x}$Mn$_2$O$_7$ has revealed that the corresponding energy splitting is about 0.4 eV.[19] The XAS measurements of the layered nickelate La$_{2-x}$Sr$_x$NiO$_4$ also show an $e_g$ orbital splitting of about 0.7 eV.[20] This implies that the electron-lattice couplings of the 4$d$ and 5$d$ orbitals are stronger than those of the 3$d$ orbitals due to their extended nature.

We will now discuss the low energy spectral features observed in $\sigma(\omega)$ below 1.5 eV. As shown in Figs. 2(a) and (b), the $\sigma(\omega)$ of $Sr_2RuO_4$ and $Sr_2RhO_4$ exhibit Drude-like peaks, which are the characteristic optical response of a metallic state. The spectral weight of the Drude-like peak for $Sr_2RhO_4$ is one-third that of $Sr_2RuO_4$, which might be associated with the less metallic character of $Sr_2RhO_4$. Note that the Ru-O-Ru bond angle of $Sr_2RuO_4$ is 180°. In $Sr_2RhO_4$, the $RhO_6$ octahedra are rotated with respect to the $c$-axis, so the $d$ electron bands of $Sr_2RhO_4$ should become narrower. In addition, the number of $d$ electrons increases, so the electron correlation should increase.

In contrast to the metallic electrodynamics in $Sr_2RuO_4$ and $Sr_2RhO_4$, the $\sigma(\omega)$ of $Sr_2IrO_4$ shows an insulating behavior with an optical gap around 0.3 eV. The sharp spikes below 0.1 eV are due to optical phonon modes. Interestingly, $Sr_2IrO_4$ has a two-peak structure below 2 eV with a sharp peak near 0.5 eV. These spectral features below 2 eV are attributable to the splitting of the $t_{2g}$ bands into the subbands below and above the $E_F$, as displayed schematically in Fig. 1(c). This splitting will lead to the $d$-$d$ transition below the charge transfer excitations, as shown in Fig. 2(c).

However, the origin of the two-peak structure of $Sr_2IrO_4$ has not been determined. One possibility is the structural distortion in the electronic structure of $Sr_2IrO_4$. According to the recent angle-resolved photoemission spectroscopy result for $Sr_2RhO_4$, the mixing between the $d\,xy$ and $d\,x^2\text{-}y^2$ bands driven by the rotation of the $RhO_6$ octahedra could lead to an almost fully occupied $d\,xy$ band with the remaining three electrons in $d\,yz/zx$ bands,[21] and the $d\,yz/zx$ bands in the layered structure are likely to have a one-dimensional nature.[22] Due to the structural similarity of $Sr_2RhO_4$ and $Sr_2IrO_4$, we postulate that similar effects hold for $Sr_2IrO_4$. Then, the quasi-one-dimensional $d\,yz/zx$ bands might be subject to density wave instability, which could induce the insulating state of $Sr_2IrO_4$ and the fairly sharp peak in the $\sigma(\omega)$ near 0.5 eV. Indeed, Cao *et al*. suggested the possibility of a charge density wave in $Sr_2IrO_4$ based on its negative differential resistivity behavior.[10] However, the amount of structural distortion, *i.e.*, the rotation of the metal-oxygen octahedra, is nearly the same in $Sr_2RhO_4$ and $Sr_2IrO_4$. Further systematic studies of the low energy peaks and a structural analysis of $Sr_2IrO_4$ are needed.

In summary, we investigated the electronic structures of $Sr_2MO_4$ (*M*=Ru, Rh, and Ir) systematically by measuring optical and XAS spectra. These spectroscopic studies demonstrated that these three compounds have similar interband transitions, indicating that their electronic structures are quite similar. The $e_g$ orbital splittings were found to be very large, suggesting interesting roles of electron-lattice coupling.

We thank J. S. Lee for valuable discussions. Experiments at PLS were supported in part by the MOST and POSTECH. This work was financially supported Creative Research Initiatives (Functionally Integrated Oxide Heterostructure) of MOST/KOSEF, eSSC at POSTECH, and BK21. YSL was supported by the Soongsil University Research Fund.

**References**

[1] J. G. Bednorz and K. A. Muller, Z. Phys. B **64**, 189 (1986).

[2] J. M. Tranquada, B. J. Sternlieb, J. D. Axe, Y. Nakamura, and S. Uchida, Nature (London), **375**, 561 (1995).

[3] Y. Tokura and N. Nagaosa, Science, **288**, 462 (2000).

[4] Y. Maeno, H. Hashimoto, K. Yoshida, S. Nishizaki, T. Fujita, J. G. Bednorz, and F. Lichtenberg, Nature, **372**, 532 (1994).

[5] Y. Moritomo, T. Arima, and Y. Tokura, J. Phys. Soc. Jpn. **64**, 4117 (1995).

[6] J. H. Jung, K. H. Kim, D. J. Eom, T. W. Noh, E. J. Choi, J. Yu, Y. S. Kwon, and Y. Chung, Phys. Rev. B **55**, 15489 (1997).

[7] H.-J. Noh, S.-J. Oh, B.-G. Park, J.-H. Park, J.-Y. Kim, H.-D. Kim, T. Mizokawa, L. H. Tjeng, H.-J. Lin, C. T. Chen, S. Schuppler, S. Nakatsuji, H. Fukazawa, and Y. Maeno, Phys. Rev. B **72**, 052411 (2005).

[8] O. Friedt, M. Braden, G. Andre, P. Adelmann, S. Nakatsuji, and Y. Maeno, Phys. Rev. B **63**, 174432 (2001).

[9] M. Itoh, T. Shimura, Y. Inaguma, and Y. Morii. J. Solid State Chem. **118**, 206 (1995).

[10] G. Cao, J Bolivar, S. McCall, J. E. Crow, and R. P. Guertin, Phys. Rev. B **57**, R11039 (1998).

[11] I. Nagai and S. I. Ikeda, *to be published*.

[12] S. Nakatsuji and Y. Maeno, Phys. Rev. B **62**, 6458 (2000).

[13] J. S. Lee, Y. S. Lee, T. W. Noh, K. Char, J. Park, S.-J. Oh, J.-H. Park, C. B. Eom, T. Takeda, and R. Kanno, Phys. Rev. B **64**, 245107 (2001).

[14] Y. S. Lee, J. S. Lee, T. W. Noh, D. Y. Byun, K. S. Yoo, K. Yamamura, and E. Takayama-Muromachi, Phys. Rev. B **67**, 113101 (2003).

[15] T. Arima, Y. Tokura, and J. B. Torrance, Phys. Rev. B **48**, 17006 (1993).

[16] J. Matsuno, Y. Okimoto, M. Kawasaki, and Y. Tokura, Phys. Rev. Lett. **95**, 176404 (2005).

[17] J. S. Lee, Y. S. Lee, T. W. Noh, S.-J. Oh, J. Yu, S. Nakatsuji, H. Fukazawa, and Y. Maeno, Phys. Rev. Lett. **89**, 257402 (2002).

[18] Z. Fang, N. Nagaosa, and K. Terakura, Phys. Rev. Lett. **69**, 045116 (2004).

[19] J.-H. Park, T. Kimura, and Y. Tokura, Phys. Rev. B **58**, R13330 (1998).

[20] P. Kuiper, J. van Elp, D. E. Rice, D. J. Buttrey, H.-J. Lin, and C. T. Chen, Phys. Rev. B **57**, 1552 (1998).

[21] B. J. Kim *et al.*, *in press*.

[22] A. Damascelli, D. H. Lu, K. M. Shen, N. P. Armitage, F. Ronning, D. L. Feng, C. Kim, Z.-X. Shen, T. Kimura, Y. Tokura, Z. Q. Mao, and Y. Maeno, Phys. Rev. Lett.

**85**, 5194 (2000).

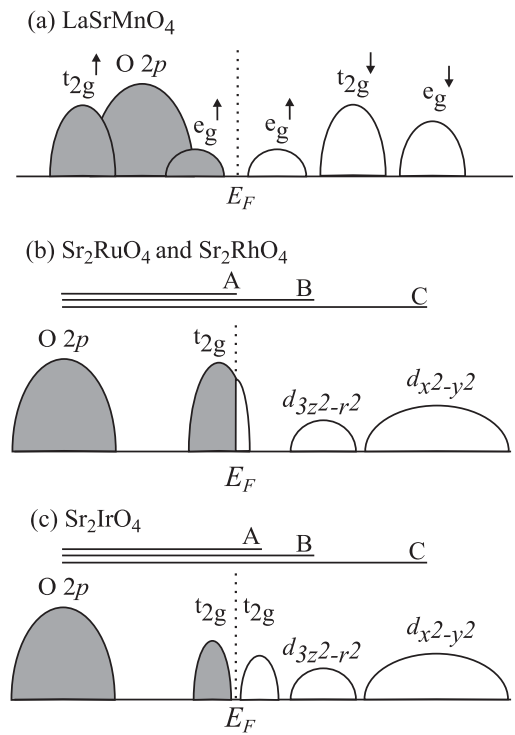

Fig. 1. Schematic diagrams of the electronic structures of (a) LaSrMnO$_4$, (b) Sr$_2$RuO$_4$ and Sr$_2$RhO$_4$, and (c) Sr$_2$IrO$_4$. $E_F$ represents the Fermi level. The arrows pointing up and down indicate spin-up and spin-down, respectively. The schematic diagram of the electronic structure of LaSrMnO$_4$ is drawn based on Ref. 5.

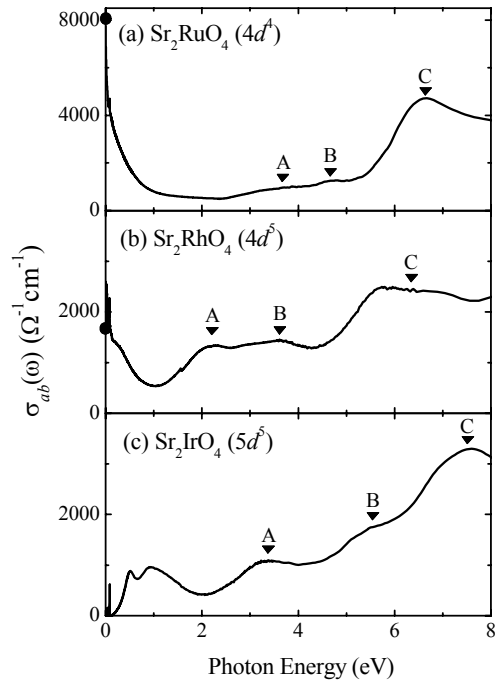

Fig. 2. The *ab*-plane $\sigma(\omega)$ of (a) $Sr_2RuO_4$, (b) $Sr_2RhO_4$, and (c) $Sr_2IrO_4$ at room temperature. The solid triangles and letters represent the optical transitions, which are shown in Fig. 1. The solid circles are the dc conductivities of $Sr_2RuO_4$ and $Sr_2RhO_4$ at room temperature.[10,11]

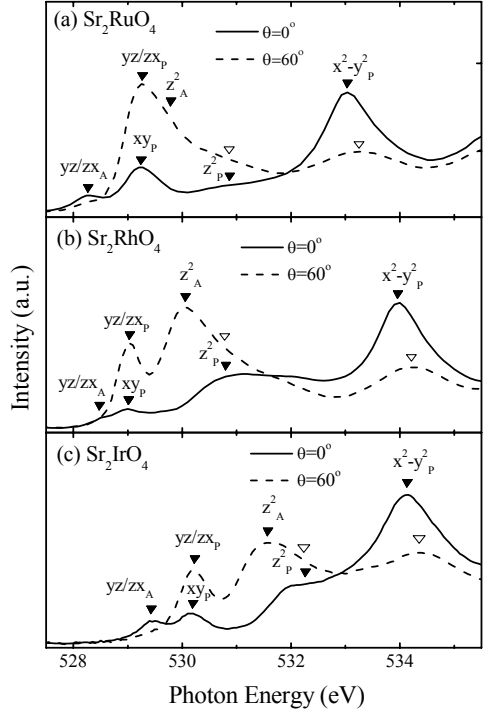

Fig. 3. Polarization-dependent O 1$s$ XAS spectra of (a) $Sr_2RuO_4$, (b) $Sr_2RhO_4$, and (c) $Sr_2IrO_4$. The spectra of $Sr_2RuO_4$ are taken from Ref. 7. $\theta$ is the incidence angle of light to the surface normal. The solid triangles and labels represent the positions and characters of empty $d$ bands, respectively. The open triangles indicate $z^2_p$ and $x^2-y^2_p$ states, which are shown in the $\theta$=60° spectrum due to the partial in-plane polarization of incident light.

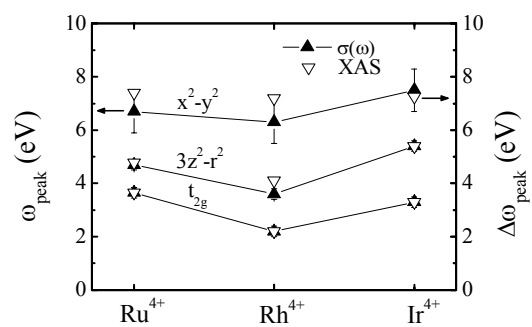

Fig. 4. The peak positions of the interband transitions in $\sigma(\omega)$ and relative peak positions of the XAS spectra of $Sr_2MO_4$. In order to compare two spectra, we shifted the energy values of the empty $t_{2g}$ states in the XAS spectra to the values of the corresponding optical transitions: O $2p \rightarrow M\ t_{2g}$.